\documentclass[a4paper,conference]{IEEEtran}
\usepackage[left=1.57cm,right=1.57cm,top=0.95cm,bottom=2.54cm]{geometry}
\IEEEoverridecommandlockouts
\usepackage{amsmath,amssymb,amsfonts}
\usepackage{algorithmic}
\usepackage{graphicx}
\usepackage{textcomp}
\usepackage{xcolor}
\usepackage[backend=biber]{biblatex}
\usepackage{tikz}
\usepackage{circuitikz}
\usepackage{hyperref}
\usepackage{flushend}
\usepackage{pgfplots}
\usepackage{breakurl}
\pgfplotsset{compat=1.18}
\begin{document}

\title{\vspace*{1cm}Boosting Cross-Architectural Emulation Performance by Foregoing the Intermediate Representation Model}

\author{\IEEEauthorblockN{Amy I. Parker\thanks{This work has been submitted to the IEEE for possible publication. Copyright may be transferred without notice, after which this version may not longer be accessible.}}
\IEEEauthorblockA{\textit{Department of Computer Science} \\
\textit{California State University, Fullerton}\\
Fullerton, CA, USA \\
amyipdev@csu.fullerton.edu}
}

\maketitle

\begin{abstract}
As more applications utilize virtualization and emulation to run mission-critical tasks, the performance requirements of emulated and virtualized platforms continue to rise. Hardware virtualization is not universally available for all systems, and is incapable of emulating CPU architectures, requiring software emulation to be used. QEMU, the premier cross-architecture emulator for Linux and some BSD systems, currently uses dynamic binary translation (DBT) through intermediate representations using its Tiny Code Generator (TCG) model. While using intermediate representations of translated code allows QEMU to quickly add new host and guest architectures, it creates additional steps in the emulation pipeline which decrease performance. We construct a proof of concept emulator to demonstrate the slowdown caused by the usage of intermediate representations in TCG; this emulator performed up to 35x faster than QEMU with TCG, indicating substantial room for improvement in QEMU's design. We propose an expansion of QEMU's two-tier engine system (Linux KVM versus TCG) to include a middle tier using direct binary translation for commonly paired architectures such as RISC-V, x86, and ARM. This approach provides a slidable trade-off between development effort and performance depending on the needs of end users.
\end{abstract}


\section{Introduction}
Cross-architectural emulation is a vital tool for many applications, from transitioning between CPU architectures \cite{10.5555/1855741.1855754} to ensuring backward compatibility in enterprise system upgrades \cite{10533089} and providing cross-platform software development and testing with systems more powerful than the target device. End users utilize cross-architectural emulation for these purposes, as well as other more niche purposes such as digital preservation of applications \cite{qemudigprev}. These emulators allow software compiled for a different instruction set architecture from the host to execute using either execution simulation or through binary translation, which result in the original executable producing equivalent results to if it had been compiled for the host system's architecture. Effective cross-architecture emulators allow executables to be essentially portable within the same operating system, as long as all libraries are either statically linked or have versions for the guest architecture present on new hosts.

Virtualization and emulation are vital to modern technology stacks, as virtual machines have become the industry standard for high-security application contexts \cite{10.1007/978-3-642-54525-2_8}. As a result, the performance impacts of virtualization and emulation have been studied extensively. In particular, the performance of technologies like Xen and KVM --- which leverage hardware extensions provided by CPUs to substantially increase execution speed to near-native levels --- has been the subject of much research \cite{5708625} \cite{7367280} \cite{6714189}, as they do not require substantial binary translation. Although these technologies have recently expanded to more platforms and architectures \cite{7367280}, they are not usable for cross-architectural emulation, as these hardware extensions simply allow host instructions to be executed in a virtually separate environment of the same CPU architecture \cite{6714189}. In other words, these technologies can only run programs designed for the same architecture; they cannot be used to emulate across architectures.

On Linux and several BSD platforms, the current most developed cross-architectural emulation platform is the Quick Emulator (QEMU), which serves as the effective base implementation for KVM on Linux \cite{5708625}. QEMU allows a user to either emulate an entire system or run processes from binaries compiled for any of approximately 30 CPU architectures. When KVM is unavailable, either due to the host lacking hardware support or the host and guest architectures being different, QEMU switches to a software emulator called the Tiny Code Generator (TCG), which translates guest instructions first into an intermediate representation and then into host instructions.

This middle step --- the conversion to an intermediate representation (IR) --- imbues a substantial performance penalty compared to direct binary translation. While the intermediate representation step allows for a simplified instruction optimization pipeline and for the creation of an $N \times N$ architectural mapping (all architectures can emulate all architectures) with only $2N$ implementations (a TCG frontend and a TCG backend), common architecture pairings --- such as \texttt{x86\_64} and \texttt{riscv64}, or \texttt{aarch64} and \texttt{x86\_64} --- suffer unnecessary performance penalties and lose out on potential optimizations specific to the particular architecture pairing. Traditional emulators have not employed the IR model, leading to a lack of existing research on TCG's effect on performance.

Our contributions in this paper are:

\begin{enumerate}
    \item We profiled QEMU, and in particular the TCG module, to examine how the intermediate representation step in TCG's binary translation works and the performance impact it has.
    \item We developed a proof-of-concept emulator framework for the RISC-V architecture (64-bit, base instruction set only) that behaves equivalently to QEMU's user mode \texttt{qemu-riscv64} emulator.
    \item We created synthetic benchmarks for testing the emulators' performances in scenarios that can accurately account for the difference between direct instruction emulation and QEMU's binary translation.
    \item Based on the results of the benchmarks, we proposed a new three-tier system for QEMU that can incorporate high-usage architecture pairings through direct translation rather than being limited to IR-based TCG and KVM.
\end{enumerate}

The paper is organized as follows. \S\ref{sec:prelim} provides background information on the architecture of QEMU, the TCG code generator, and the RISC-V instruction set architecture. \S\ref{sec:ps} provides a working definition of the problem addressed by this paper. \S\ref{sec:design} details the design of the proof-of-concept emulator. \S\ref{sec:eval} describes the designed benchmarks, provides their results, and discusses them. \S\ref{sec:prior} reviews related works in the field. Finally, \S\ref{sec:end} concludes.


\section{Preliminaries} \label{sec:prelim}
\subsection{QEMU Architecture}

QEMU has two main categories of user front-ends: system-mode interfaces and user-mode interfaces. System-mode interfaces emulate a full computer system, including devices and the boot process; this mode is typically used for virtual machines and testing systems software in a realistic environment. User-mode interfaces provide process-level emulation, running programs for another architecture under the host's kernel as if it were programmed for the host's architecture. User mode uses guest-native libraries, loading them as standard shared libraries and executing the instructions rather than the kernel. Both modes support the same overall instruction set architectures, with system-mode having more tunables to match the exact architecture of individual CPUs. While this research is applicable to both system-mode and user-mode, user-mode is easier to implement and test, and is used for this research.

Unlike system-mode, user-mode does not support KVM. While this research is focused on \textit{cross}-architecture emulation, it should be noted that if a user were to run same-architecture applications through QEMU user-mode, they would suffer equivalent performance hits to a user running a cross-architecture application, despite the necessary binary translation being incredibly minimal. 

Each architecture supported by QEMU generates two executable binaries: \texttt{qemu-ARCH} and \texttt{qemu-system-ARCH}, where \texttt{ARCH} is the instruction set architecture. \texttt{qemu-ARCH} is the user-mode emulator, and runs as a standalone executable that does not link to an overall QEMU library. Upon invocation with a path pointing to an existing executable, QEMU loads the ELF data from the file (if valid) and begins constructing a parallel memory model. Original ELF sections need to remain in memory in the event of an irregular pointer access, and QEMU constructs a memory graph accordingly \cite{qemumemory}. Additional memory regions have direct host-compatible instructions which are executed by the host, either by the KVM engine or by the host directly after TCG conversion. 

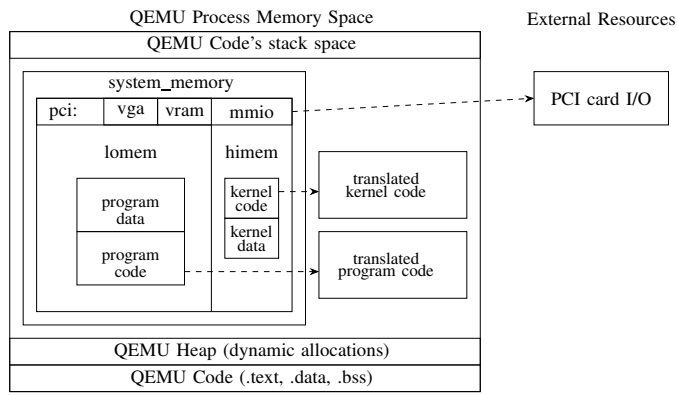
\begin{figure}[!ht]
\centering
\resizebox{0.5\textwidth}{!}{%
\begin{circuitikz}
\tikzstyle{every node}=[font=\small]
\draw  (2.75,15.5) rectangle (11.5,8.75);
\node [font=\normalsize] at (7.25,15.75) {QEMU Process Memory Space};
\draw  (2.75,8.75) rectangle (11.5,9.25);
\node [font=\normalsize] at (7.25,9) {QEMU Code (.text, .data, .bss)};
\draw  (2.75,15.5) rectangle (11.5,15);
\node [font=\normalsize] at (7.25,15.25) {QEMU Code's stack space};
\node [font=\normalsize] at (13.75,15.75) {External Resources};
\draw  (2.75,9.75) rectangle (11.5,9.25);
\node [font=\normalsize] at (7.25,9.5) {QEMU Heap (dynamic allocations)};
\node [font=\normalsize] at (5.75,14.5) {system\_memory};
\draw  (3.25,14.25) rectangle (8,10.25);
\draw  (3,14.75) rectangle (8.25,10);
\draw  (8.5,13.25) rectangle (11.25,12);
\draw  (8.5,11.75) rectangle (11.25,10.5);
\draw  (3.25,14.25) rectangle (8,13.75);
\node [font=\normalsize] at (3.75,14) {pci:};
\draw  (4.5,14.25) rectangle (5.5,13.75);
\draw  (5.5,14.25) rectangle (6.5,13.75);
\node [font=\normalsize] at (5,14) {vga};
\node [font=\normalsize] at (6,14) {vram};
\node [font=\normalsize] at (7.25,14) {mmio};
\draw  (12.5,14.75) rectangle (15,13.75);
\node [font=\normalsize] at (13.75,14.25) {PCI card I/O};
\draw [->, >=Stealth, dashed] (8,14) -- (12.5,14.25);
\draw [short] (6.5,13.75) -- (6.5,10.25);
\node [font=\normalsize] at (5,13.25) {lomem};
\node [font=\normalsize] at (7.25,13.25) {himem};
\draw  (6.75,12.75) rectangle (7.75,12);
\node [font=\small] at (7.25,12.5) {kernel};
\node [font=\small] at (7.25,12.25) {code};
\draw  (6.75,12) rectangle (7.75,11.25);
\node [font=\small] at (7.25,11.5) {data};
\node [font=\small] at (7.25,11.75) {kernel};
\node [font=\small] at (9.75,12.75) {translated};
\node [font=\small] at (9.75,12.5) {kernel code};
\draw  (4,12.75) rectangle (6,11.75);
\draw  (4,11.75) rectangle (6,10.75);
\node [font=\small] at (9.75,11.25) {translated};
\node [font=\small] at (9.75,11) {program code};
\node [font=\small] at (5,12.25) {program};
\node [font=\small] at (5,12) {data};
\node [font=\small] at (5,11.25) {program};
\node [font=\small] at (5,11) {code};
\draw [->, >=Stealth, dashed] (6,11) -- (8.5,11);
\draw [->, >=Stealth, dashed] (7.75,12.5) -- (8.5,12.5);
\end{circuitikz}
}%
\caption{QEMU memory architecture based on the official API example \cite{qemumemory}}
\label{fig:qemu_memarch}
\end{figure}

Once QEMU has initialized the memory, it passes execution over to the applicable engine to begin execution. For TCG, QEMU traps exceptions, interrupts, and system calls to ensure that it handles them itself due to differences in exception numbering and behavior between architectures; for instance, the \texttt{write()} system call is number 64 for 64-bit Linux RISC-V systems, while \texttt{x86\_64} Linux systems use number 1.

\subsection{TCG Architecture}

\begin{figure*}[!ht]
\centering
\resizebox{1\textwidth}{!}{%
\begin{circuitikz}
\tikzstyle{every node}=[font=\small]
\draw  (2.5,31) -- (4.375,29.875) -- (6.25,31) -- (4.375,32.125) -- cycle;
\node [font=\small] at (4.375,31.125) {PC already seen?};
\draw  (7.5,31.5) rectangle  node {\small Fetch} (9.5,30.75);
\draw  (10.5,31.5) rectangle  node {\small Decode} (12.5,30.75);
\draw  (5.75,29.25) rectangle (7.75,29);
\draw  (5.75,29) rectangle  node {\scriptsize Instruction} (7.75,28.75);
\draw  (5.75,28.75) rectangle (7.75,28.5);
\draw  (5.75,28.5) rectangle (7.75,28.25);
\draw  (10,30) rectangle  node {\small TCG ops buffer} (13,29);
\draw  (13.5,31) -- (15.125,30.125) -- (16.75,31) -- (15.125,31.875) -- cycle;
\node [font=\small] at (15.125,31.125) {Branch?};
\draw  (14.25,29.25) rectangle  node {\small TCG} (16.25,28.5);
\draw  (17.5,31.5) rectangle  node {\small Execute} (19.5,30.75);
\draw  (17.5,29.25) rectangle (19.5,29);
\draw  (17.5,29) rectangle (19.5,28.75);
\draw  (17.5,28.75) rectangle  node {\scriptsize Translation cache} (19.5,28.5);
\draw  (17.5,28.5) rectangle (19.5,28.25);
\draw  (11.5,28) ellipse (1.25cm and 0.5cm) node {\small Built-in ops} ;
\draw [short] (2.5,31) -- (1.25,31);
\draw [short] (1.25,31) -- (1.25,33.75);
\draw [short] (1.25,33.75) -- (19,33.75);
\draw [->, >=Stealth] (19,33.75) -- (19,31.5);
\draw [short] (18.25,31.5) -- (18.25,33.5);
\draw [short] (18.25,33.5) -- (4.25,33.5);
\draw [->, >=Stealth] (4.25,33.5) -- (4.25,32.25);
\draw [->, >=Stealth] (6.25,31.25) -- (7.5,31.25);
\draw [->, >=Stealth] (9.5,31.25) -- (10.5,31.25);
\draw [->, >=Stealth] (12.5,31.25) -- (13.75,31.25);
\draw [short] (15.25,32) -- (15.25,32.5);
\draw [short] (15.25,32.5) -- (8.5,32.5);
\draw [->, >=Stealth] (8.5,32.5) -- (8.5,31.5);
\draw [->, >=Stealth] (15,30.25) -- (15,29.25);
\draw [short] (16.25,29.25) -- (17,29.25);
\draw [short] (17,29.25) -- (17,30.75);
\draw [->, >=Stealth] (17,30.75) -- (17.5,30.75);
\draw [->, >=Stealth, dashed] (13,29) -- (14.25,29);
\draw [->, >=Stealth, dashed] (11.5,28.5) -- (14.25,28.5);
\draw [->, >=Stealth, dashed] (16.25,28.5) -- (17.5,28.5);
\draw [->, >=Stealth, dashed] (18.5,29.25) -- (18.5,30.75);
\draw [dashed] (7.75,29) -- (8.5,29);
\draw [->, >=Stealth, dashed] (8.5,29) -- (8.5,30.75);
\draw [->, >=Stealth, dashed] (11.5,30.75) -- (11.5,29.75);
\draw [ dashed] (6.5,33.25) rectangle  (16.75,30.25);
\draw [ dashed] (13.5,30) rectangle  (16.75,27.25);
\node [font=\small] at (15.25,27.5) {Code Generation};
\node [font=\small] at (8,32.75) {Binary Translation};
\node [font=\small] at (7,31) {No};
\node [font=\small] at (2.25,30.75) {Yes};
\node [font=\small] at (15.75,32) {No};
\node [font=\small] at (15.5,29.5) {Yes};
\node [font=\small] at (18.5,28) {Host instruction cache};
\node [font=\small] at (6.75,28) {Source instructions (ELF)};
\node [font=\small] at (11.5,27.25) {Pre-translated instructions};
\end{circuitikz}
}%
\caption{Overall TCG execution model, adapted from Gligor et al. \cite{10.1145/1629435.1629446}}
\label{fig:tcg_bigpicture}
\end{figure*}
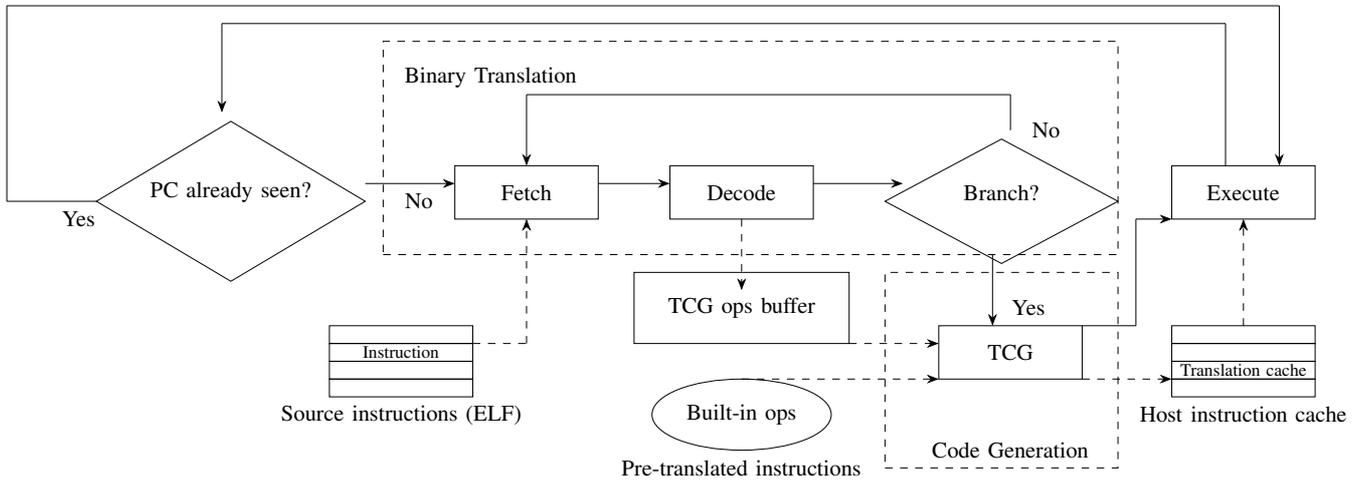

Once TCG takes control of execution, instructions at previously unseen memory addresses are run through the binary translation pipeline. Fig.\, \ref{fig:tcg_bigpicture} demonstrates the overall flow of instructions through this pipeline, where untranslated addresses are checked through the instruction cache and, if the translated instruction is not present, sent through TCG's IR-based translation model.

When instructions need to run through TCG, they first go through a TCG frontend, which translates the instructions from the guest architecture to \texttt{TCGOp}s, the IR system for TCG \cite{qemu_tcgir}. \texttt{TCGOp}s then run through an optimization pipeline that folds constant expressions, compresses scalar instructions to vector instructions on supported platforms, and removes unreachable code \cite{qemu_gitlab_opt}. After optimization, the cleaned \texttt{TCGOp}s are sent to the TCG backend for the host architecture, where they are finally translated into the instructions that will be loaded into the memory and executed.

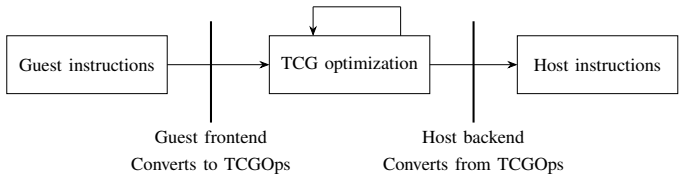
\begin{figure}[!ht]
\centering
\resizebox{0.5\textwidth}{!}{%
\begin{circuitikz}
\tikzstyle{every node}=[font=\small]
\draw  (1.75,29.5) rectangle  node {\small Guest instructions} (4.5,28.5);
\draw  (6.25,29.5) rectangle  node {\small TCG optimization} (9,28.5);
\draw  (10.5,29.5) rectangle  node {\small Host instructions} (13.25,28.5);
\draw [->, >=Stealth] (4.5,29) -- (6.25,29);
\draw [->, >=Stealth] (9,29) -- (10.5,29);
\draw [line width=1pt, short] (5.25,29.75) -- (5.25,28);
\draw [line width=1pt, short] (9.75,29.75) -- (9.75,28);
\draw [short] (8.5,29.5) -- (8.5,30);
\draw [short] (8.5,30) -- (7,30);
\draw [->, >=Stealth] (7,30) -- (7,29.5);
\node [font=\small] at (5.25,27.75) {Guest frontend};
\node [font=\small] at (5.25,27.25) {Converts to TCGOps};
\node [font=\small] at (9.75,27.75) {Host backend};
\node [font=\small] at (9.75,27.25) {Converts from TCGOps};
\end{circuitikz}
}%
\caption{TCG IR pipeline}
\label{fig:tcg_ir}
\end{figure}

\subsection{RISC-V ISA}

RISC-V is an open, royalty-free, reduced instruction set computing (RISC) architecture that has gained prominence due to its freely accessible specifications \cite{riscvspecsite} and the reasonable availability of development tools and build toolchains for the architecture \cite{10.1145/3603781.3603942} relative to other open RISC architectures. The architecture has base standards for 32-bit, 64-bit, and 128-bit CPUs, and a wide variety of extensions to support floating-point arithmetic, hardware multiplication, atomic operations, and other instructions. This flexibility has positioned it as one of the most successful new CPU architectures; it competes effectively with ARM in the low-end embedded microprocessor market, and adoption is projected to grow at $25\%$ per year \cite{jasr_riscv}.

While RISC-V can be used in a regular operating system environment, it has its roots in embedded computing devices, which makes it ideal for developing proof-of-concept solutions. Tooling a RISC-V-targeting compiler to avoid relocations and the C runtime --- which add hefty amounts of development time to a proof-of-concept emulator --- is relatively trivial in comparison to compilers for x86 and ARM. RISC-V can also be easily reduced down to the lowest common denominator for a given bit size --- the base integer extensions --- without conflicting with tooling for various microarchitectures like on ARM.

\section{Problem Statement} \label{sec:ps}

For hosts that do not support hardware virtualization through KVM, who are using QEMU user mode, or who are running guest code of a different architecture, QEMU only provides TCG as an engine choice. The current performance of QEMU TCG is not reasonable for running large-scale, high-performance, or low-latency applications, such as full operating systems, data processing tools, or equipment operational tools. If the performance was closer to native or to KVM performance, then using these tools with TCG would be reasonable; however, it is currently unreasonable to operate over TCG for these applications. Not all architecture pairings need faster emulation --- it is highly unlikely that an \texttt{m68k} user would need to run high-speed \texttt{alpha} code, for instance --- but some do; a solution is necessary which responsibly utilizes developer time (as the TCG IR model does) while providing performance improvements to often used architecture pairings.

\section{Design} \label{sec:design}

We developed \texttt{riscv-um}, a RISC-V 64-bit proof of concept emulator built in Rust.\footnote{The source code for \texttt{riscv-um} is available at \url{https://github.com/amyipdev/riscv-um} under the GNU General Public License, version 2.} The emulator supports a limited subset of the base integer instruction set, only containing instructions that were actually used within the benchmarks; the remaining instructions can be implemented with a minimal performance penalty, as the only increase in time would be a few microseconds while loading the larger emulator executable into memory. 

The emulator does not perform direct binary translation with instructions; instead, it executes them using a register array and a memory interface. As a proof of concept, a direct binary translator would only be necessary if the performance differences between this emulator and QEMU on low-branching benchmarks are minimal; as noted later, we did not observe this in our results. A simulator will be faster in low-branch environments, but not high-branch ones, as there is no instruction caching or direct execution of previously translated instructions; however, for the purposes of comparing the lengths of the translation pipeline, a simulator is more than sufficient when the difference is of a sufficient magnitude.

After loading in the binary through a similar method to QEMU and setting up its memory structure --- which is simpler than QEMU's memory model, as it creates a fixed memory space with simple address translation rather than having a full tree of memory regions --- \texttt{riscv-um} begins a regular fetch-decode-execute cycle, with results being written back as appropriate to the register file (a shared \texttt{\&mut [u64; 32]}) and the memory. Instructions are first decoded using a per-opcode jump table, with sub-functions handled by \texttt{match} statements; it is up to the optimizing compiler whether the \texttt{match} blocks are treated as jump tables or as a series of connected conditionals.

\begin{figure}[!ht]
\centering
\resizebox{0.5\textwidth}{!}{%
\begin{circuitikz}
\tikzstyle{every node}=[font=\small]
\draw  (4,13) rectangle  node {\normalsize Fetch} (6,11.75);
\draw  (7.75,13) rectangle  node {\normalsize Decode op} (9.75,11.75);
\draw  (10.75,11.5) rectangle  node {\normalsize Decode fct} (12.75,10.25);
\draw  (13.5,13) rectangle  node {\normalsize Execute} (15.5,11.75);
\draw [->, >=Stealth] (6,12.5) -- (7.75,12.5);
\draw [->, >=Stealth] (9.75,12.5) -- (13.5,12.5);
\draw [short] (9,11.75) -- (9,10.75);
\draw [->, >=Stealth] (9,10.75) -- (10.75,10.75);
\draw [short] (12.75,10.75) -- (14.5,10.75);
\draw [->, >=Stealth] (14.5,10.75) -- (14.5,11.75);
\node [font=\normalsize] at (11.5,12.75) {if no fct in type};
\node [font=\normalsize] at (9.5,11) {else};
\draw  (4.25,16.25) rectangle (9,14.5);
\draw [short] (7.25,16.25) -- (7.25,14.5);
\node [font=\normalsize] at (6.5,16.5) {mem (*mut c\_void)};
\node [font=\normalsize] at (5.75,15.25) {.text,.bss};
\node [font=\normalsize] at (8.25,15.25) {stack};
\draw  (11.5,16.25) rectangle  node {\footnotesize 0} (11.75,16);
\node [font=\normalsize] at (12.25,15.5) {};
\draw  (11.5,15.25) rectangle (11.75,15);
\draw  (11.5,15) rectangle (11.75,14.75);
\draw  (11.5,14.5) rectangle (11.75,14.25);
\draw  (11.5,15.75) rectangle (11.75,15.5);
\draw  (11.5,15.5) rectangle (11.75,15.25);
\draw  (11.5,14.75) rectangle (11.75,14.5);
\draw  (11.5,16) rectangle (11.75,15.75);
\draw  (11.75,16.25) rectangle (12,16);
\draw  (11.75,15.25) rectangle (12,15);
\draw  (11.75,15) rectangle (12,14.75);
\draw  (11.75,14.5) rectangle (12,14.25);
\draw  (11.75,15.75) rectangle (12,15.5);
\draw  (11.75,15.5) rectangle (12,15.25);
\draw  (11.75,14.75) rectangle (12,14.5);
\draw  (11.75,16) rectangle (12,15.75);
\node [font=\normalsize] at (12.25,15.5) {};
\draw  (12.25,15.5) rectangle (12.5,15.25);
\draw  (12.25,16) rectangle (12.5,15.75);
\node [font=\normalsize] at (12.25,16) {};
\draw  (12.25,16) rectangle (12.5,15.75);
\draw  (12.25,16.25) rectangle (12.5,16);
\draw  (12.25,15.75) rectangle (12.5,15.5);
\draw  (12.25,15) rectangle (12.5,14.75);
\draw  (12.25,15.25) rectangle (12.5,15);
\draw  (12.25,14.75) rectangle (12.5,14.5);
\draw  (12.25,14.5) rectangle (12.5,14.25);
\node [font=\normalsize] at (12,15.5) {};
\draw  (12,15.5) rectangle (12.25,15.25);
\draw  (12,16) rectangle (12.25,15.75);
\node [font=\normalsize] at (12,16) {};
\draw  (12,16) rectangle (12.25,15.75);
\draw  (12,16.25) rectangle (12.25,16);
\draw  (12,15.75) rectangle (12.25,15.5);
\draw  (12,15) rectangle (12.25,14.75);
\draw  (12,15.25) rectangle (12.25,15);
\draw  (12,14.75) rectangle (12.25,14.5);
\draw  (12,14.5) rectangle (12.25,14.25);
\draw  (14.75,15.75) rectangle  node {\normalsize pc (\&mut u64)} (17.75,15);
\node [font=\normalsize] at (12,16.5) {registers (\&mut [u64; 32])};
\draw  (16,13.5) -- (17.25,12.75) -- (18.5,13.5) -- (17.25,14.25) -- cycle;
\node [font=\normalsize] at (17.25,13.5) {branch?};
\draw [->, >=Stealth] (15.5,13) -- (17,13);
\draw [->, >=Stealth] (16,13.5) -- (16,15);
\draw [->, >=Stealth] (17.25,14.25) -- (17.25,15);
\node [font=\small] at (15,14) {branch addr};
\node [font=\small] at (17.75,14.75) {yes};
\node [font=\small] at (15.5,14.25) {no};
\node [font=\small] at (17.75,14.5) {pc+4};
\draw [short] (13.5,13) -- (13.5,14.25);
\draw [->, >=Stealth] (13.5,14.25) -- (12.5,14.25);
\draw [short] (13.5,13) -- (13.25,13);
\draw [short] (13.25,13) -- (13.25,14);
\draw [short] (13.25,14) -- (7.25,14);
\draw [->, >=Stealth] (7.25,14) -- (7.25,14.5);
\draw [->, >=Stealth] (4.5,14.5) -- (4.5,13);
\draw [short] (7,14.5) -- (7,13.75);
\draw [short] (7,13.75) -- (13.75,13.75);
\draw [->, >=Stealth] (13.75,13.75) -- (13.75,13);
\draw [short] (7.75,14.5) -- (7.75,13.5);
\draw [short] (7.75,13.5) -- (14,13.5);
\draw [->, >=Stealth] (14,13.5) -- (14,13);
\draw [short] (12.5,14.5) -- (13.75,14.5);
\draw [short] (13.75,14.5) -- (13.75,14);
\draw [short] (13.75,14) -- (14,14);
\draw [short] (14,14) -- (14,13.75);
\draw [short] (14,13.75) -- (14.25,13.75);
\draw [->, >=Stealth] (14.25,13.75) -- (14.25,13);
\draw [short] (14.75,15.75) -- (14.75,17);
\draw [short] (14.75,17) -- (4,17);
\draw [->, >=Stealth] (4,17) -- (4,13);
\draw  (16.25,11.5) rectangle  node {\small syscalls} (18,10.75);
\draw [short] (15.5,12) -- (17.25,12);
\draw [->, >=Stealth] (17.25,12) -- (17.25,11.5);
\draw [short] (6.25,14.5) -- (6.25,10);
\draw [short] (6.25,10) -- (17,10);
\draw [->, >=Stealth] (17,10) -- (17,10.75);
\draw [short] (17.25,10.75) -- (17.25,9.75);
\draw [short] (17.25,9.75) -- (6.5,9.75);
\draw [->, >=Stealth] (6.5,9.75) -- (6.5,14.5);
\end{circuitikz}
}%
\caption{\texttt{riscv-um} execution model and data pathway}
\label{fig:riscvum_model}
\end{figure}
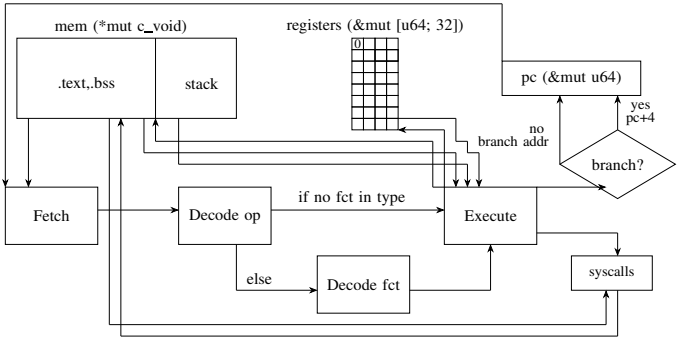

As this emulator is a proof-of-concept and will not be used in production, several liberties were taken which minimally improve performance (mostly through avoiding additional checks and processing) at the risk of potential vulnerabilities. For instance, memory access violations are handled with a pass-through model, where invalid memory accesses based on the address translation would throw a segmentation fault that is passed onto the user; this method could allow for emulator data itself to be read or even altered. These issues do not apply to the implementation of the proposed design in QEMU, as QEMU's memory API is not susceptible to these basic attacks.

To implement this in QEMU, we propose altering the decision structure QEMU uses in system-mode when choosing the engine by adding a third option. If KVM is unavailable, but the architecture pairing is in a set of implemented pairings (for instance, \texttt{x86\_64} host and \texttt{riscv64} guest), QEMU launches a different engine that, while utilizing the overall QEMU architecture, has a binary translation model designed specifically for the two architectures in the pairing. If such a pairing does not exist, QEMU can fall back on TCG for emulation. To implement this in the user-mode, a similar decision structure can be added to the user-mode executables, running before the current main functions (which currently always launch TCG).

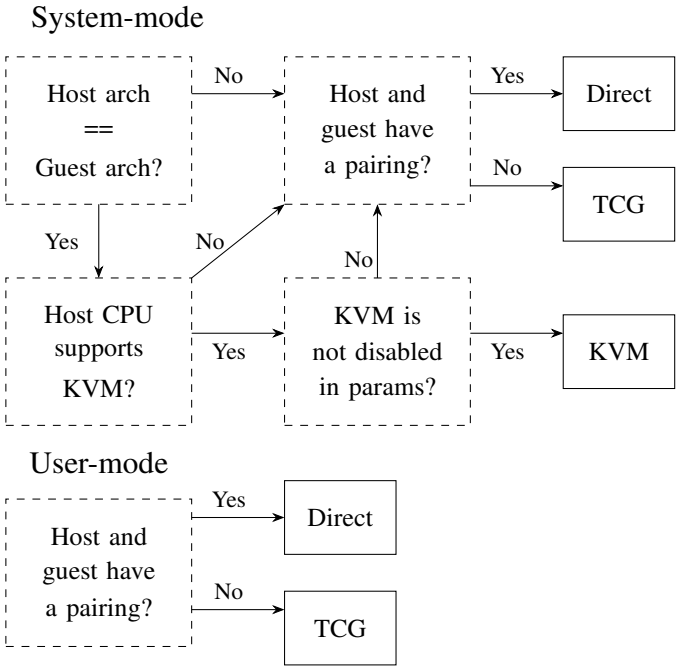
\begin{figure}[!ht]
\centering
\resizebox{0.5\textwidth}{!}{%
\begin{circuitikz}
\tikzstyle{every node}=[font=\normalsize]
\node [font=\large] at (3.5,16.5) {System-mode};
\node [font=\large] at (3.25,10.5) {User-mode};
\draw [ dashed] (2,16) rectangle  (4.5,14);
\node [font=\normalsize] at (3.25,15.5) {Host arch};
\node [font=\small] at (2.75,13.5) {Yes};
\node [font=\normalsize] at (3.25,15) {==};
\node [font=\normalsize] at (3.25,14.5) {Guest arch?};
\draw [ dashed] (2,13) rectangle  (4.5,11);
\node [font=\normalsize] at (3.25,12.5) {Host CPU};
\node [font=\normalsize] at (3.25,12) {supports};
\node [font=\normalsize] at (3.25,11.5) {KVM?};
\draw [->, >=Stealth] (3.25,14) -- (3.25,13);
\draw [ dashed] (5.75,13) rectangle  (8.25,11);
\node [font=\normalsize] at (7,12.5) {KVM is};
\node [font=\normalsize] at (7,12) {not disabled};
\node [font=\normalsize] at (7,11.5) {in params?};
\draw [->, >=Stealth] (4.5,12.25) -- (5.75,12.25);
\node [font=\small] at (8.75,15.75) {Yes};
\node [font=\small] at (8.75,12) {Yes};
\node [font=\small] at (5,12) {Yes};
\node [font=\small] at (4.75,13.5) {No};
\node [font=\small] at (5,15.75) {No};
\node [font=\small] at (8.75,14.5) {No};
\node [font=\small] at (6.75,13.25) {No};
\draw  (9.5,12.5) rectangle  node {\normalsize KVM} (11,11.5);
\draw [->, >=Stealth] (8.25,12.25) -- (9.5,12.25);
\draw [ dashed] (5.75,16) rectangle  (8.25,14);
\node [font=\normalsize] at (7,15.5) {Host and};
\node [font=\normalsize] at (7,15) {guest have};
\node [font=\normalsize] at (7,14.5) {a pairing?};
\draw  (9.5,16) rectangle  node {\normalsize Direct} (11,15);
\draw  (9.5,14.5) rectangle  node {\normalsize TCG} (11,13.5);
\draw [->, >=Stealth] (4.5,15.5) -- (5.75,15.5);
\draw [->, >=Stealth] (8.25,15.5) -- (9.5,15.5);
\draw [->, >=Stealth] (8.25,14.25) -- (9.5,14.25);
\draw [->, >=Stealth] (7,13) -- (7,14);
\draw [->, >=Stealth] (4.5,13) -- (5.75,14);
\draw [ dashed] (2,10) rectangle  (4.5,8);
\node [font=\normalsize] at (3.25,9.5) {Host and};
\node [font=\normalsize] at (3.25,9) {guest have};
\node [font=\normalsize] at (3.25,8.5) {a pairing?};
\node [font=\small] at (5,10) {Yes};
\node [font=\small] at (5,8.75) {No};
\draw  (5.75,10.25) rectangle  node {\normalsize Direct} (7.25,9.25);
\draw  (5.75,8.75) rectangle  node {\normalsize TCG} (7.25,7.75);
\draw [->, >=Stealth] (4.5,9.75) -- (5.75,9.75);
\draw [->, >=Stealth] (4.5,8.5) -- (5.75,8.5);
\end{circuitikz}
}%
\caption{Proposed QEMU engine choice structure}
\label{fig:qemu_choice}
\end{figure}

\section{Evaluation} \label{sec:eval}

To test the performance of the emulators and evaluate how much additional overhead TCG brings to the execution environment, we developed a benchmark called \texttt{benchgen} which outputs a customizable number of instructions. The benchmark outputs a rotation of \texttt{add}, \texttt{sub}, and \texttt{sll} instructions with the registers used in the instructions rotating with a different period ($p = 3$ for instructions, $p = 4$ for registers). 

With a selection of 2 million instructions, the registers were initialized at \texttt{t0} $= 8745425$, \texttt{t1} $= 2413112$, \texttt{t2} $= 51124341$, and \texttt{t3} $= 991232131$, which were randomly entered. The correct outputs were precomputed on the host system using a separate script as \texttt{t0} $= 8697740129876948287$, \texttt{t1} $= 0$, \texttt{t2} = $= 9749003943832603329$, and \texttt{t3} $= 18220595702735330224$.

This model was chosen because, unlike traditional benchmarks, this test does not have a significant amount of branch instructions. QEMU uses translation caching and does not retranslate instructions at a given address, which gives it a substantial advantage over any simulation-based emulator, but not over a direct translation emulator. As such, using a traditional benchmark would not accurately reflect the contribution of \textit{TCG} to runtimes; a simple N-Queens benchmark shows this, where with $n=27$ QEMU completed the test in $0.799$ seconds while \texttt{riscv-um} completed it in $7.103$ seconds on the same hardware.

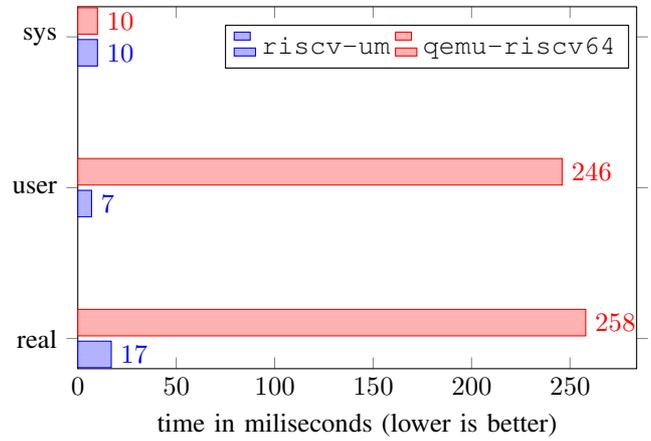
\begin{figure}
    \centering
    \begin{tikzpicture}
        \begin{axis}[xbar, symbolic y coords={real,user,sys}, ytick=data, xlabel={time in miliseconds (lower is better)}, width=.5\textwidth, legend style={at={(0.625,0.95)}, anchor=north, legend columns=2}, nodes near coords, xmin=0, y=2cm]
            \addplot table[x=rvum,y=stat]{\thedata};
            \addplot table[x=qemu,y=stat]{\thedata};
            \legend{\texttt{riscv-um}, \texttt{qemu-riscv64}}
        \end{axis}
    \end{tikzpicture}
    \caption{\texttt{riscv-um} vs QEMU results}
    \label{fig:theresults}
\end{figure}

Both benchmarks were run on a Framework 16 with a Ryzen 7 7840HS running NixOS Unstable while plugged in to consistent A/C power. For \texttt{riscv-um}, the command run was \texttt{time ../target/release/riscv-um ./rb}, and for QEMU \texttt{time qemu-riscv64 ./rb}; the \texttt{time} implementation was from GNU Bash 5.2.37. The same packages and benchmark binaries were used for both tests, ensured through the usage of the same Nix shell environment when running both tests. While the results shown in Fig.\, \ref{fig:theresults} are for a single pass of the benchmark, later runs showed roughly the same results, and the difference in magnitude between \texttt{riscv-um} and QEMU was roughly equal across tested hosts.

\pgfplotstableread[row sep=\\, col sep=&]{
    stat & rvum & qemu \\
    real & 17 & 258 \\
    user & 7 & 246 \\ 
    sys & 10 & 10 \\
}\thedata

While the system times were the same, indicating that loading the RISC-V binaries and returning took the same amount of time for both emulators, \texttt{riscv-um} far outperformed QEMU overall, completing the actual computation in $7$ ms compared to QEMU's $246$ ms, which is a $ \frac{246}{7} = 35 \frac{1}{7} \approx 35\times $ performance improvement over QEMU. 

These results are unlikely to be obtained to such magnitude in real emulation, as binary translation is more computationally intensive than device simulation. However, as long as a binary translation interpretation takes \textit{less than $35 \times$} the amount of work as \texttt{riscv-um}'s simulation, it will beat QEMU and provide a performance improvement. Assuming a reasonable cost multiplier of $2\mbox{--}3.5\times$, this would mean between a $ 10\times = 1000\%$ and $17.5 \times = 1750\%$ performance benefit, which solves the fundamental problem QEMU faces for cross-architectural applications.

\section{Related Work} \label{sec:prior}

Dung et al. did a significant amount of work profiling and mapping QEMU's architecture for instruction execution when using TCG while developing methods for simulating alternative caching structures and measuring their latency \cite{6945730}. Their profiling work built off of basic profiling work done by Gligor et al., who were researching simulation of multiprocessor SoCs \cite{10.1145/1629435.1629446}. While most of their research touches on QEMU's handling of caches, the portions on QEMU's overall architecture are very informative and detailed. In particular, they offer one of the most complete big-picture overviews of the TCG engine in the current literature. However, their research does not cover the more in-depth parts of the TCG engine that are relevant to this research; mentions of the IR system are completely absent, as they are obscured under the \texttt{tcg\_gen\_code()} function due to their work occurring outside of the code generation module. 

Michel et al. conducted research on optimizing outputted SIMD instructions by adding additional signals from the original guest instructions through the TCG engine to the architecture backends \cite{qemusimd}. By adding additional architecture-specific SIMD hints (particularly targeting ARM Neon architecture for guest code) to the IR system yielded up to a $400\%$ speed improvement. While this was just a single set of instructions added to the TCG IR, it demonstrates that optimizations based specifically on the guest architecture --- and potentially on both the guest and host architectures --- have potential that is not being realized currently by TCG. Although it does not directly bypass the IR step in TCG, it does provide evidence that better performance through direct translation is possible.

Fu et al. also researched SIMD optimization in QEMU, and did so by creating a custom variant of the TCG IR pipeline with 32-bit ARM and x86 frontends and an \texttt{x86\_64} backend \cite{7092577}. This work bypassed QEMU's helper functions and created new vector IR instructions that could be downcasted to scalar instructions if necessary, similar to the work done by Michel et al. Fu's research observes many inefficiencies in the TCG system, and achieves better results (up to $7.6 \times$) through optimizations targeting specific architecture pairings --- but still uses the IR model, leaving it very inefficient.

Luo et al. developed a variant of QEMU that simulates out-of-order processors' operations correctly per CPU cycle, and in doing so investigated the structure of TCG \cite{6429075}. In addition to further researching the helper functions bypassed by Fu et al., Luo et al. investigated further the loss of context that occurs as instructions enter the TCG IR pipeline. Properly simulating the cycles of out-of-order processors required additional extractions from the QEMU engine at the frontend prior to entering the IR pipeline, with the backend later receiving the information and using it to simulate complex behaviors through the helper functions. 

\section{Conclusions and Future Work} \label{sec:end}

QEMU provides the backbone of modern cross-architectural emulation for a majority of platforms, and is a critical tool for end-users, developers, and industries alike. Dynamic binary translation (DBT) in QEMU is an effective overall architecture for emulation, but the usage of an intermediate representation (IR)-based translation model introduces unacceptable performance penalties for many applications. Bypassing TCG and conducting direct emulation, particularly through direct binary translation, has the potential to substantially increase performance for high-importance architecture pairings. Our proof-of-concept model was able to demonstrate a theoretical $35 \times$ performance increase from leaving behind TCG's intermediate representation system and implementing direct translation for specific architecture pairs.

By adding a third engine to QEMU's arsenal, performance for key architecture pairings in both system-mode and user-mode can be substantially increased. A challenge in doing so will be refactoring the QEMU codebase to develop a cohesive model for direct binary translation, utilizing both the existing non-translational resources in TCG and allowing for the full potential of architecture-pair-specific optimizations to be realized. Future work should focus on creating effective, maintainable implementations within the QEMU codebase, working to eventually reach the automatic selection model described in Fig.\, \ref{fig:qemu_choice}.

\section*{Acknowledgment}

I thank my research advisor, Dr. Mikhail I. Gofman, for the incredible assistance and guidance he has provided in bringing this paper to fruition. His expertise and advice has been critical throughout the research for this paper.

\printbibliography

\end{document}